\documentstyle [12pt]{article}

\begin{document}

\begin{titlepage}

\title{Ladder proof of nonlocality without inequalities 
and without probabilities}

\author{Ad\'{a}n Cabello\thanks{Electronic address: fite1z1@sis.ucm.es}\\
{\em Departamento de F\'{\i}sica Aplicada,}\\
{\em Universidad de Sevilla, 41012 Sevilla, Spain.}}

%First version: 29 October, 1997
%This version: 
\date{\today}

\maketitle

\begin{abstract}
The ladder proof of nonlocality without inequalities for 
two spin-$\frac{1}{2}$ particles
proposed by Hardy \cite{Hardy97} and Hardy et al.\ \cite{BBDH97} works 
only for nonmaximally entangled states and goes through for 
$50\%$ of pairs at the most. A similar ladder proof for 
two spin-$1$ particles in a 
maximally entangled state is presented. In its simplest
form, the proof goes through for $17\%$ of pairs.
An extended version works for $100\%$ of pairs. The
proof can be extended to any maximally entangled state of two
spin-$s$ particles (with $s\ge 1$).

\end{abstract}

\end{titlepage}

\section{Introduction}
Recently, Hardy \cite{Hardy97} and Hardy et al.\ \cite{BBDH97} have 
presented a generalization of Hardy's proof of Bell's theorem without inequalities 
for two spin-$\frac{1}{2}$ particles \cite{Hardy93}. 
Neither the original proof
nor the ladder generalization and
not even the generalizations 
for the case of two spin-$s$ particles proposed in \cite{CN92,GK97}
work for maximally entangled states. 
The improvement in the ladder generalization
of the two spin-$\frac{1}{2}$ particles case 
comes from the fact that, adding new observables, 
the probability for the proof to go through 
increases.
It grows from 9\% of pairs in the original proof with two 
alternative observables on each particle 
\cite{Hardy93} to almost 50\% of pairs when an infinite number 
of alternative observables are considered \cite{Hardy97,BBDH97}. 

In this paper, I present a similar ladder proof of Bell's theorem without
inequalities for two spin-$1$ particles prepared in the singlet state. 
In its simplest form, treated in section~3, the proof works for 17\% of pairs. 
An extended version, considered in 
section~4, works for 100\% of pairs and uses a finite number of alternative
observables. In section~5, I explain how the proof can be extended to
any other maximally entangled state of two spin-$1$ particles
and to any maximally entangled state of two spin-$s$ particles (with $s\ge 1$).
The advantages and disadvantages of the proposed ladder proof
in order to design a real experiment to test Bell's theorem are discussed
in section~6. 
Finally, in section~7, some differences between 
ladder proofs of Bell's theorem and the proofs of the so-called
Kochen-Specker with locality theorem initially proposed by Heywood and
Redhead \cite{HR83} are remarked upon. 
In order to introduce some notations, I begin in section~2 with a 
brief review of the ladder proof by Hardy et al. 

\section{Ladder proof for two spin-$\frac{1}{2}$ particles}
The scenario considered by Hardy et al.\ \cite{Hardy97,BBDH97} is the following:
Two spin-$\frac{1}{2}$ particles, initially prepared in some 
specific quantum state, are confined to space-like 
separated regions of space-time. 
On the first particle only one measurement $\hat A_k$ chosen from
a set $\{\hat A_j\}_{j=0}^K$ can be made.
Each one of these potential measurements has the outcome $A_k$ or 
$A_k^\bot$. Similarly, on the second particle only one measurement 
$\hat B_k$ from the set $\{\hat B_j\}_{j=0}^K$ can be made to see whether it has 
the outcome $B_k$ or $B_k^\bot$. 
Hardy et al.\ show that there exist quantum states 
$\left| {\eta} \right\rangle $ and sets of measurements $\{\hat A_j\}_{j=0}^K$ and
$\{\hat B_j\}_{j=0}^K$ with the following properties:
\begin{equation}
P_\eta (A_K,B_K)=P_K\ne 0\,,
\label{Hardy1}
\end{equation}
\begin{equation}
P_\eta (B_{j-1}|A_j)=1\,,
\label{Hardy2}
\end{equation}
\begin{equation}
P_\eta (A_{j-1}|B_j)=1\,,
\label{Hardy3}
\end{equation}
\begin{equation}
P_\eta (A_0,B_0)=0\,,
\label{Hardy4}
\end{equation}
for $j=1$ to $K$. From these properties we can build a ladder of 
inferences based on EPR condition for 
elements of reality ({\em ``If, without in any way disturbing a system, we can
predict with certainty (i.\ e., with a probability equal to unity) the value of a
physical quantity, then there exist an element of physical reality corresponding
to this physical quantity''} \cite{EPR35}). This ladder of
inferences will conclude in
a contradiction. We start from the proportion, $P_K$, 
of pairs in which we get $A_K$ and 
$B_K$, given by equation (\ref{Hardy1}). Then, using EPR condition,
we have the following inferences: since, according to 
(\ref{Hardy2}) ((\ref{Hardy3})), 
$A_K$ ($B_K$) allows us to 
predict with certainty $B_{K-1}$ ($A_{K-1}$), then $B_{K-1}$ ($A_{K-1}$) was 
an element of reality. Similarly, since $A_{K-1}$ ($B_{K-1}$) allows us to 
predict with certainty $B_{K-2}$ ($A_{K-2}$), then $B_{K-2}$ ($A_{K-2}$) was 
an element of reality, etc. 
Finally,
since $A_1$ allows us to predict $B_0$ and $B_1$ allows us to predict $A_0$,
$B_0$ and $A_0$ were both elements of reality. 
Therefore, in the state $\left| {\eta} \right\rangle $ 
we should have $A_0$ and $B_0$ for at least the proportion $P_K$ of pairs. 
But, according to (\ref{Hardy4}), we should never get $A_0$ and $B_0$. 
So we reach 
a contradiction. In fact, this way of 
viewing the contradiction is not unique \cite{Hardy97}.

\begin{sloppypar}
The whole reasoning can be summarized with the aid of some graphs. 
The graph 
\begin{picture}(90,15)
\put(30,0){\circle*{5}}
\put(60,0){\circle*{5}}
\put(30,0){\line(1,0){30}}
\put(6,0){$A_K$}
\put(69,0){$B_K$}
\put(38,5){$P_K$}
\end{picture}
represents the statement (\ref{Hardy1}): the outcomes $A_K$ and $B_K$ occur 
together with probability $P_K$. 
Graphs like
\begin{picture}(100,10)
\put(60,0){\circle*{5}}
\put(30,0){\circle*{5}}
\put(30,0){\line(1,0){30}}
\put(30,0){\vector(1,0){18}}
\put(69,0){$B_{j-1}$}
\put(6,0){$A_j$}
\end{picture}
represent statements like (\ref{Hardy2}): if $A_j$ happens then we can
predict $B_{j-1}$ with certainty. Analogously, graphs like
\begin{picture}(100,10)
\put(36,0){\circle*{5}}
\put(66,0){\circle*{5}}
\put(66,0){\line(-1,0){30}}
\put(66,0){\vector(-1,0){18}}
\put(4,0){$A_{j-1}$}
\put(75,0){$B_j$}
\end{picture}
represent statements like (\ref{Hardy3}). 
Finally, the graph
\begin{picture}(90,10)
\put(30,0){\circle*{5}}
\put(60,0){\circle*{5}}
\put(30,0){\line(1,0){30}}
\put(44,4){\line(0,-1){8}}
\put(46,4){\line(0,-1){8}}
\put(8,0){$A_0$}
\put(69,0){$B_0$}
\end{picture}
represents the statement (\ref{Hardy4}): the outcomes $A_0$ and $B_0$ never
occur together. Using these graphs, the 
ladder proofs by Hardy et al.\ with two, three and $K+1$ 
settings are represented in Figure~1. For the case of two alternative 
observables on each particle, the maximum value of
$P_1$ is 9.0\% \cite{Hardy93}. Adding more observables this value grows.
In case of three observables on each particle, the maximum value of 
$P_2$ is 17.5\% \cite{Hardy97,BBDH97}. As $K$ tends to infinity 
$P_{K}$ tends to 50\% \cite{Hardy97,BBDH97}. For details on these 
calculations the reader is referred
to \cite{Hardy97,BBDH97,Hardy93}. The name ``ladder'' comes from the fact that
the proof uses a chain---of adjustable length---of predictions with certainty.
\end{sloppypar}

On the other hand, Clifton and Niemann \cite{CN92} 
and, recently, Ghosh and Kar \cite{GK97} have proposed 
generalizations for the case of two spin-$s$ particles 
(with $s\ge\frac{1}{2}$) of Hardy's original proof. Their 
generalizations are based on 
statements similar to (\ref{Hardy1})-(\ref{Hardy4}) in which 
all the measurements $\hat A_j$ and 
$\hat B_j$ ($j=0,1$) correspond 
to nondegenerate operators (components of spin).
In case of two spin-$1$ particles, Ghosh and Kar have found a 
maximum value for $P_1$ of 13.2\% \cite{GK97}. 
Both Clifton and Niemann's extension 
\cite{CN92} and Ghosh and Kar's extension \cite{GK97} do not work
for maximally entangled states.

\section{Ladder proof for two spin-$1$ particles in the singlet state}
As far as I know, no ladder proof for maximally entangled states
exists. However, the Hilbert space corresponding to a
system of two spin-$s$ particles, 
${\cal H}_{2s+1} \otimes {\cal H}_{2s+1}$, with $s\ge 1$, is richer than
the Hilbert space 
${\cal H}_2 \otimes {\cal H}_2$ of two spin-$\frac{1}{2}$ particles. 
In particular, if $s\ge 1$,
we can measure and predict
the outcomes of local observables corresponding to degenerated operators. 
For instance, in case of 
two spin-$1$ particles, there are sets of three 
mutually compatible local observables 
which can be measured on the same run of the experiment. 
In this paper I exploit these facts to construct a 
ladder proof without inequalities of Bell's theorem for maximally 
entangled states of two spin-$1$ particles.

The scenario is analogous to the one described in section~2, changing 
the two space-like separated spin-$\frac{1}{2}$ 
particles prepared in a nonmaximally
entangled state by two space-like separated spin-$1$ 
particles prepared in the singlet
state 
\begin{equation}
\left| \psi \right\rangle =
{1 \over {\sqrt 3}} 
\left({\left| \hbar \right\rangle \left| {-\hbar} \right\rangle +
\left| {-\hbar} \right\rangle \left| \hbar \right\rangle -
\left| 0 \right\rangle \left| 0 \right\rangle } \right)\,.
\label{singlet}
\end{equation}
On the first particle a measurement of the {\em square} of the spin component
in some direction ${\bf n}_k$, $({\bf S}_1\cdot {\bf n}_k)^2$, chosen from 
a large but specific set of them can be made. Each one of these possible 
measurements has the outcome $0$ or $\hbar ^2$. However, in the singlet state,
\begin{equation}
P_\psi \left({[{\bf S}_2\cdot {\bf n}_k]^2=0 
\left|{\,[{\bf S}_1\cdot {\bf n}_k]^2=0
} \right.} \right)=1\,,
\label{uno}
\end{equation}
\begin{equation}
P_\psi \left({[{\bf S}_2\cdot {\bf n}_k]^2=\hbar ^2 \left|{\,[{\bf S}_1\cdot 
{\bf n}_k]^2=\hbar ^2} \right.} \right)=1\,,
\label{dos}
\end{equation}
and therefore, if the outcome of $({\bf S}_1\cdot {\bf n}_k)^2$ 
is $0$ ($\hbar ^2$), we can predict 
with certainty that
the outcome of $({\bf S}_2\cdot {\bf n}_k)^2$ will be $0$ ($\hbar ^2$).
Properties (\ref{uno}) and (\ref{dos}), and the corresponding properties 
obtained by interchanging 
particle 1 and 2, will be used in our proof in the same way as properties 
(\ref{Hardy2}) and (\ref{Hardy3}) in the ladder proof for two spin-$\frac{1}{2}$
particles. To summarize this kind of inferences I shall continue to 
use the same kind of graphs as in the previous proof.

In addition, in case of two spin-$1$ particles in
the singlet state, if the outcome of measuring 
 $({\bf S}_1\cdot {\bf n}_k)^2$ is $0$, then we can predict with certainty that
the outcome of measuring $({\bf S}_2\cdot {\bf n}_j)^2$, 
in every direction ${\bf n}_j$ 
orthogonal to the direction ${\bf n}_k$, will be $\hbar ^2$. This occurs because
\begin{equation}
P_\psi \left({[{\bf S}_2\cdot {\bf n}_j]^2=
\hbar ^2 \left|{\,[{\bf S}_1\cdot {\bf n}_k]^2=0
} \right.} \right)=1\,,\,\,\,\,\,\,
\forall\,{\bf n}_j \bot {\bf n}_k\,,
\label{tres}
\end{equation}
and, consequently $({\bf S}_1\cdot {\bf n}_j)^2$ and 
$({\bf S}_2\cdot {\bf n}_k)^2$ cannot be both zero if ${\bf n}_j$ is 
orthogonal to ${\bf n}_k$, i.\ e.,
\begin{equation}
P_\psi \left([{\bf S}_1\cdot {\bf n}_j]^2=0,[{\bf S}_2\cdot {\bf n}_k]^2=0 
\right)=0\,,\,\,\,\,\,\,
\forall\,{\bf n}_j \bot {\bf n}_k\,.
\label{tresdos}
\end{equation}
Property (\ref{tres}) can be used to predict with certainty the result of 
more than one measurement
on the second particle from a single measurement on the first (or vice versa). 
New graphs must be
introduced to reflect these new inferences. For instance, the graph
\begin{picture}(90,10)
\put(30,0){\circle*{5}}
\put(6,0){$A_k$}
\put(60,7.5){\circle*{5}}
\put(30,0){\line(4,1){30}}
\put(30,0){\vector(4,1){18}}
\put(69,7.5){$B_l$}
\put(60,-7.5){\circle*{5}}
\put(30,0){\line(4,-1){30}}
\put(30,0){\vector(4,-1){18}}
\put(69,-7.5){$B_m$}
\end{picture}
represents the case of two predictions: 
$A_k$ could be $({\bf S}_1\cdot {\bf n}_k)^2=0$
and $B_l$ and $B_m$ could be $({\bf S}_2\cdot {\bf n}_l)^2=\hbar ^2$ and 
$({\bf S}_2\cdot {\bf n}_m)^2=\hbar ^2$, respectively, 
being ${\bf n}_l$ and ${\bf n}_m$ 
both orthogonal to ${\bf n}_k$.
Similarly, by interchanging the first and the 
second particles there are also graphs like
\begin{picture}(90,10)
\put(60,0){\circle*{5}}
\put(69,0){$B_k$}
\put(30,-7.5){\circle*{5}}
\put(60,0){\line(-4,-1){30}}
\put(60,0){\vector(-4,-1){18}}
\put(6,7.5){$A_l$}
\put(30,7.5){\circle*{5}}
\put(60,0){\line(-4,1){30}}
\put(60,0){\vector(-4,1){18}}
\put(6,-7.5){$A_m$}
\end{picture}.
Property (\ref{tresdos}) plays the same role as property (\ref{Hardy4}) in the
ladder proof by Hardy et al.,
therefore, to represent it I will use the same graph as in section~2.

Moreover, for a spin-$1$ particle the observables
$({\bf S}_1\cdot {\bf n}_i)^2$, $({\bf S}_1\cdot {\bf n}_j)^2$, 
$({\bf S}_1\cdot {\bf n}_k)^2$ are compatible if ${\bf n}_i$, ${\bf n}_j$, 
${\bf n}_k$ are mutually orthogonal directions. In fact, their values sum
$2\hbar ^2$
\begin{equation}
({\bf S}_1\cdot {\bf n}_i)^2 +({\bf S}_1\cdot {\bf n}_j)^2+
({\bf S}_1\cdot {\bf n}_k)^2=2 \hbar ^2\,.
\label{resolution}
\end{equation}
Therefore, in the singlet state,
if the outcome of measuring $({\bf S}_1\cdot {\bf n}_i)^2$ is $\hbar ^2$
{\em and} the outcome of measuring $({\bf S}_1\cdot {\bf n}_j)^2$ 
is $\hbar ^2$ then, using (\ref{uno}), we can predict with certainty that
the outcome of measuring $({\bf S}_2\cdot {\bf n}_k)^2$ will be $0$,
\begin{equation}
P_\psi \left( {[{\bf S}_2\cdot{\bf n}_k]^2=0 \left| 
{\,[{\bf S}_1\cdot {\bf n}_i]^2=\hbar ^2\,\,\&\,\,
[{\bf S}_1\cdot {\bf n}_j]^2=\hbar ^2} \right.} \right)=1\,.
\label{cuatro}
\end{equation}
To represent these inferences I will use a new kind of graph:
\begin{picture}(90,10)
\put(60,0){\circle*{5}}
\put(69,0){$B_k$}
\put(30,7.5){\circle*{5}}
\put(6,7.5){$A_i$}
\put(30,-7.5){\circle*{5}}
\put(6,-7.5){$A_j$}
\thicklines
\put(30,7,5){\line(4,-1){30}}
\put(30,7.5){\vector(4,-1){18}}
\put(30,-7.5){\line(4,1){30}}
\put(30,-7.5){\vector(4,1){18}}
\end{picture}. 
Or, interchanging the particles, 
\begin{picture}(90,10)
\put(30,0){\circle*{5}}
\put(6,0){$A_k$}
\put(60,7.5){\circle*{5}}
\put(69,7.5){$B_i$}
\put(60,-7.5){\circle*{5}}
\put(69,-7.5){$B_j$}
\thicklines
\put(60,7.5){\line(-4,-1){30}}
\put(60,7.5){\vector(-4,-1){18}}
\put(60,-7.5){\line(-4,1){30}}
\put(60,-7.5){\vector(-4,1){18}}
\end{picture}.
To avoid confusion when different kinds of graph appear, note that 
the latter have thicker lines than the previous graphs.

\subsection{First part: stepladder argument}
The proof itself has two parts. In the first part, using a
chain of predictions with certainty, I will show that no
local realistic interpretation exists for the case in which the
outcome of measuring $({\bf S}_1\cdot {\bf i})^2$ on the first particle is $0$ 
and the outcome of measuring $({\bf S}_2\cdot {\bf a})^2$
on the second particle is also $0$ when the directions ${\bf i}$ and
${\bf a}$ form an angle $\phi$ bound between certain values. This chain of 
predictions is summarized in Figure~2 and will be explicitly explained in
the following.

\begin{sloppypar}
Let $A_4$ be the outcome $[{\bf S}_1\cdot (1 ,0 ,0)]^2=0$ and
let $B_4$ be the outcome 
$[{\bf S}_2\cdot (\cos \phi ,\sin \phi ,0)]^2=0$.
In the singlet state, 
\begin{equation}
P_\psi (A_4,B_4)=P_4=\frac{1}{3}\cos ^2\phi\,.
\label{once}
\end{equation}
Thus, if $\phi$ is not $\frac{\pi}{2}$, the probability $P_4$ is not zero.
Let $A_3$ be 
$[{\bf S}_1\cdot (\tan \phi ,-1 , \cot \theta)]^2=\hbar ^2$ and let $A_2$ be
$[{\bf S}_1\cdot (\tan \phi ,-1 , -\cot \theta)]^2=\hbar ^2$,
where $\theta$ is not $\frac{\pi}{2}$). Then, 
in the singlet state,
\begin{equation}
P_\psi (A_3\,|\,B_4)=1\,,
\label{one}
\end{equation}
\begin{equation}
P_\psi (A_2\,|\,B_4)=1\,.
\label{two}
\end{equation}
Analogously, let $B_3$ be 
$[{\bf S}_2\cdot (0 ,\cos \theta ,-\sin \theta)]^2=\hbar ^2$ and let $B_2$ be
$[{\bf S}_2\cdot (0 ,\cos \theta ,\sin \theta)]^2=\hbar ^2$. Then 
\begin{equation}
P_\psi (B_3\,|\,A_4)=1\,,
\label{three}
\end{equation}
\begin{equation}
P_\psi (B_2\,|\,A_4)=1\,.
\label{four}
\end{equation}
Therefore, if $A_4$ and $B_4$ are found, then we can say that 
$A_2$, $A_3$, $B_2$ and $B_3$ were elements of reality in the sense of
EPR. 

Next, let $A_1$ be 
$[{\bf S}_1\cdot (0 ,\cos \theta ,-\sin \theta)]^2=\hbar ^2$
and let $B_1$ be
$[{\bf S}_2\cdot (\tan \phi ,-1 , \cot \theta)]^2=\hbar ^2$.
In the singlet state, 
\begin{equation}
P_\psi (A_1\,|\,B_3)=1\,,
\label{five}
\end{equation}
and
\begin{equation}
P_\psi (B_1\,|\,A_3)=1\,.
\label{six}
\end{equation}
Therefore, $A_1$ and $B_1$ were also elements of reality.
Finally, let $A_0$ be
$[{\bf S}_1\cdot (\cot \phi \csc ^2\theta ,1,-\cot \theta )]^2=0$
and let $B_0$ be
$[{\bf S}_2\cdot (\cot \phi \csc ^2\theta ,1,\cot \theta )]^2=0$.
Since in the singlet, 
\begin{equation}
P_\psi (A_0\,|\,B_1\,\,\&\,\,B_2)=1\,,
\label{seven}
\end{equation}
and
\begin{equation}
P_\psi (B_0\,|\,A_1\,\,\&\,\,A_2)=1\,,
\label{eight}
\end{equation}
then we conclude that $A_0$ and $B_0$ were elements of reality and
should be found in the singlet state at least with probability $P_4$.
However, it is easy to see that $A_0$ and $B_0$ never happen together, i.\ e.,
\begin{equation}
P_\psi (A_0,B_0)=0\,,
\label{nine}
\end{equation}
if
\begin{equation}
\cot ^2\phi=\sin ^2\theta \cos (2 \theta )\,.
\label{cinco}
\end{equation}
Since the right-hand side of (\ref{cinco}) is bound between $-1$ and $\frac{1}{8}$,
then (\ref{cinco}) is fulfilled if
\begin{equation}
\label{seis}
\arccos \left({1 \over 3} \right) \le \phi \le \arccos \left(-{1 \over 3}\right)\,,
\end{equation}
that is, if
\begin{equation}
70.5^{\circ}\le \phi \le 109.5^{\circ}\,.
\label{siete}
\end{equation}
Therefore, the maximum value of $P_4$ is
$\frac{1}{27}$.
In brief, if $\phi$ satisfies (\ref{siete}), then no local realistic description 
is possible when both $A_4$ and $B_4$ occur. At the most, $A_4$ and $B_4$ occur for
$\frac{1}{27}$ of pairs. Strictly speaking, this is not 
(yet) a ladder argument---since 
it is not apparently extensible---but just a stepladder argument.
\end{sloppypar}

\subsection{Second part: geometrical argument}
The second part uses a particular geometrical situation to improve the conclusion
of the previous stepladder argument. Let ${\bf i}$, ${\bf j}$, ${\bf k}$ be three
mutually orthogonal vectors and let ${\bf a}$, ${\bf b}$, ${\bf c}$ be other three 
mutually orthogonal vectors. Let us define
\begin{equation}
\widehat{{\bf i}\,{\bf a}}=\widehat{{\bf j}\,{\bf b}}=
\widehat{{\bf k}\,{\bf c}}=\phi_1\,,
\label{malo}
\end{equation}
\begin{equation}
\widehat{{\bf j}\,{\bf a}}=\widehat{{\bf k}\,{\bf b}}=
\widehat{{\bf i}\,{\bf c}}=\phi_2\,,
\label{bueno1}
\end{equation}
\begin{equation}
\widehat{{\bf k}\,{\bf a}}=\widehat{{\bf i}\,{\bf b}}=
\widehat{{\bf j}\,{\bf c}}=\phi_3\,.
\label{bueno2}
\end{equation}
These definitions allow us to easily
implement the orthogonality relations between the members of each triad.
These angles could be
\begin{equation}
\phi_1=\arccos \left({{1+\sqrt 3} \over 3}\right)=24.4^{\circ}\,, 
\label{phi1}
\end{equation}
\begin{equation}
\phi_2=\arccos \left({1 \over 3}\right)=70.5^{\circ}\,,
\label{phi2}
\end{equation}
\begin{equation}
\phi_3=\arccos \left({{1-\sqrt 3} \over 3}\right)=104.1^{\circ}\,.
\label{phi3}
\end{equation}
This particular situation is represented in Figure~3. 
In that case, the angles $\phi_2$, $\phi_3$ satisfy (\ref{siete}) 
but the angle $\phi_1$ does not. It is easy to see that all 9 angles 
between each direction of one triad and all three directions of the other cannot
satisfy (\ref{siete}).

Let us go back to physics. Suppose that, with the previous choice of angles,
$({\bf S}_1\cdot {\bf i})^2$, $({\bf S}_1\cdot {\bf j})^2$,
$({\bf S}_1\cdot {\bf k})^2$ are measured on the first particle 
and $({\bf S}_2\cdot {\bf a})^2$, $({\bf S}_2\cdot {\bf b})^2$, 
$({\bf S}_2\cdot {\bf c})^2$ are measured on the second. Since on each
particle we will find one $0$ and two $\hbar ^2$, then
there are 9 different possible
results. On 6 of them, the previous stepladder argument works and therefore,
in those cases, no local realistic interpretation exists. However,
the stepladder argument can be eluded when 
$({\bf S}_1\cdot {\bf i})^2=({\bf S}_2\cdot {\bf a})^2=0$, or when
$({\bf S}_1\cdot {\bf j})^2=({\bf S}_2\cdot {\bf b})^2=0$, or when
$({\bf S}_1\cdot {\bf k})^2=({\bf S}_2\cdot {\bf c})^2=0$. 
The probability for each of 
these three cases can be computed using (\ref{once}) with the election for $\phi_1$
given in (\ref{phi1}). As can be easily checked, 
the sum of these three probabilities is $0.829$. For the remaining $17.1\%$ of 
pairs the previous ladder argument goes through. In fact, it can be proved that
this is the maximum value for finding a contradiction, using definitions
(\ref{malo}-\ref{bueno2}) and if $\phi_2$ and $\phi_3$ satisfy
(\ref{siete}) and $\phi_1$ does not.

\begin{sloppypar}
\section{Ladder proof without inequalities and without probabilities}
In this section I will describe a ladder extension of the previous proof. 
Here I will
present a genuine ladder argument (i.\ e., with a variable number of steps) to
provide a proof of Bell's theorem without inequalities for two particles which
works for $100\%$ of pairs. The strategy will be the same as before. 
First I will develop the ladder argument and then I will use an additional
geometrical argument to complete the proof.

\subsection{First part: ladder argument}
The ladder argument is completely analogous to the one presented in
section~3, if longer. In fact, it would be too long to explicitly 
develop all the steps. It can be easily followed with
the aid of Figure~4. 
To simplify the diagram, in 
Figure~4 I have sometimes substituted the graph
\begin{picture}(90,10)
\put(60,0){\circle*{5}}
\put(30,0){\circle*{5}}
\put(30,0){\line(1,0){30}}
\put(30,0){\vector(1,0){18}}
\put(69,0){$B_j$}
\put(2,0){$A_{4K}$}
\end{picture}
with the graph
\begin{picture}(34,10)
\put(8,0){\circle*{5}}
\put(8,0){\circle{9}}
\put(18,0){$B_j$}
\end{picture}: i.\ e., this graph means that if $A_{4K}$ happens 
then we can predict with certainty $B_j$.
The rest of the symbols mean the same as in section~3. Please follow the
argument in Figure~4.
Note that the basic step of the ladder is composed by 4 predictions on each
particle. For instance, one basic step contains the predictions 
$A_4$ to $A_7$ and $B_4$ to $B_7$. Other basic step is the one which 
includes the predictions
$A_{4(K-2)}$ to $A_{4K-5}$ and $B_{4(K-2)}$ to $B_{4K-5}$. Note also that
the initial step (the one which contains 
$A_{4(K-1)}$ to $A_{4K}$ and $B_{4(K-1)}$ to $B_{4K}$) and the final
step (the one which contains 
$A_0$ to $A_3$ and $B_0$ to $B_3$) are both a little bit different
from the basic steps in between. The coefficients $c_j$ are
\begin{equation}
c_1=\sin \theta_1\,,
\label{coeficiente1}
\end{equation}
and, for $j\ge 2$, the coefficients can be obtained recursively using
\begin{equation}
c_{j+1}=c_{j} \cos \left( {\theta _{j+1}-\theta _j} \right)\,,
\label{coeficientej}
\end{equation}
or explicitly using 
\begin{equation}
c_j=\sin \theta _1
\prod\limits_{k=1}^{j-1} {\cos \left( {\theta _{k+1}-\theta _k} \right)}\,.
\label{coeficientej2}
\end{equation}
Therefore, $P_\psi (A_0,B_0)=0$ if
\begin{equation}
\cot ^2\phi =c_K^2\left( {\cos ^2\theta _K-\sin ^2\theta _K} \right)\,.
\label{ortogonalidad}
\end{equation}
For $K=2$, the right-hand side of (\ref{ortogonalidad}) is bound 
between $-1$ and $\left( {{{2+\sqrt {20}} \over 8}} \right)^5$, 
then (\ref{ortogonalidad}) is fulfilled if 
\begin{equation}
59.5^{\circ}\le \phi \le 120.5^{\circ}\,.
\label{K2}
\end{equation}
Analogously, for $K=3$, (\ref{ortogonalidad}) is fulfilled if 
\begin{equation}
55.2^{\circ}\le \phi \le 124.8^{\circ}\,.
\label{K3}
\end{equation}
In general, the right-hand side of (\ref{ortogonalidad}) is bound between
$-1$ and $\cos ^{2K+1}\left( {{\pi  \over {2K+1}}} \right)$. Therefore,
as $K$ tends to infinity, the right-hand side of (\ref{ortogonalidad}) is 
bound between $-1$ and $1-\epsilon$, with $\epsilon>0$, and then 
(\ref{ortogonalidad}) is fulfilled if
\begin{equation}
45^{\circ} < \phi < 135^{\circ}\,.
\label{KK}
\end{equation}
However, for our purposes, we will not need to consider an infinite
number of observables. For the following geometrical argument, 
the particular case $K=11$
will be of special interest. For $K=11$, (\ref{ortogonalidad}) 
is fulfilled if 
\begin{equation}
48.08^{\circ}\le \phi \le 131.92^{\circ}\,.
\label{K11}
\end{equation}
\end{sloppypar}

\subsection{Second part: geometrical argument}
Let us change the particular geometrical situation considered in
the second part of section~3. Maintaining the definitions 
(\ref{malo}-\ref{bueno2}), now let the relative angles between the 
mutually orthogonal vectors ${\bf i}$, ${\bf j}$,
${\bf k}$ and the mutually orthogonal vectors
${\bf a}$, ${\bf b}$, ${\bf c}$ be
\begin{equation}
\phi_1=\phi_2=\arccos \left({2 \over 3}\right)=48.19^{\circ}\,, 
\label{phi12}
\end{equation}
\begin{equation}
\phi_3=\arccos \left(-{1 \over 3}\right)=109.47^{\circ}\,.
\label{phi33}
\end{equation}
This particular situation is represented in Figure~5. With this choice of
angles, the 9 relative angles satisfy (\ref{K11}).
Therefore, whatever the results of measuring 
$({\bf S}_1\cdot {\bf i})^2$, $({\bf S}_1\cdot {\bf j})^2$,
$({\bf S}_1\cdot {\bf k})^2$ on the first particle 
and $({\bf S}_2\cdot {\bf a})^2$, $({\bf S}_2\cdot {\bf b})^2$, 
$({\bf S}_2\cdot {\bf c})^2$ on the second, we will always find the outcome
$0$ in one direction of the first particle that forms, 
with one direction of the second particle in which
the outcome $0$ has also been found, an angle $\phi$
satisfying (\ref{K11}). 
For this case, the ladder argument, when
$K=11$, gives a contradiction. Therefore, for $100\%$ of pairs 
no local realistic interpretation exists.

\begin{sloppypar}
\section{Extension to any maximally entangled state}
The above proof is based on some specific properties of the singlet state of
two spin-$1$ particles. In this section I want to argue that similar proofs exist 
for any maximally entangled state of two spin-$s$ particles (with $s\ge 1$). 
First we will see the case of two spin-$1$ particles and then the 
more general case of two spin-$s$ particles. In case of 
two spin-$1$ particles, each
maximally entangled state admits infinite Schmidt decompositions of the form
\begin{equation}
\left| \Psi  \right\rangle =
\sum\limits_{m=-1}^1 {c_m\left| {{\bf S}_1\cdot {\bf n}_j=m\hbar } \right\rangle }
\left| {{\bf S}_2\cdot {\bf n}_k=m\hbar } \right\rangle\,.
\label{Schmidt}
\end{equation}
In fact, all maximally entangled states have their nonzero Schmidt 
coefficients $c_m$ of the same absolute value \cite{EB94}
(this is true in every Schmidt
base since local unitary transformations can only change the Schmidt base vectors, 
not the Schmidt coefficients). 
On the other hand, for nonmaximally entangled states the Schmidt 
decomposition (\ref{Schmidt}) is unique.
The existence of infinite Schmidt decompositions implies that 
for every vector ${\bf n}_j$
\begin{equation}
P_\Psi \left({[{\bf S}_2\cdot {\bf n}_k]^2=0 \left|{\,[{\bf S}_1\cdot {\bf n}_j]^2=0
} \right.} \right)=1\,,
\label{unoprima}
\end{equation}
and
\begin{equation}
P_\Psi \left({[{\bf S}_2\cdot {\bf n}_k]^2=\hbar ^2 \left|{\,[{\bf S}_1\cdot 
{\bf n}_j]^2=\hbar ^2} \right.} \right)=1\,.
\label{dosprima}
\end{equation}
These two properties would play the same role in the proof as properties (\ref{uno})
and (\ref{dos}) for the singlet state. 
In addition, for a spin-$1$ particle 
$({\bf S}_2\cdot {\bf n}_k)^2$ and $({\bf S}_2\cdot {\bf n}_l)^2$ 
cannot be both zero if ${\bf n}_l$ is orthogonal to ${\bf n}_k$. 
Therefore,
$({\bf S}_1\cdot {\bf n}_j)^2$ and $({\bf S}_2\cdot {\bf n}_l)^2$ 
cannot be both zero if ${\bf n}_l$ is 
orthogonal to ${\bf n}_k$, i.\ e.,
\begin{equation}
P_\Psi \left([{\bf S}_1\cdot {\bf n}_j]^2=0,[{\bf S}_2\cdot {\bf n}_l]^2=0 
\right)=0\,,\,\,\,\,\,\,
\forall\,{\bf n}_l \bot {\bf n}_k\,,
\label{tresdosprima}
\end{equation}
and 
\begin{equation}
P_\Psi \left({[{\bf S}_2\cdot {\bf n}_l]^2=
\hbar ^2 \left|{\,[{\bf S}_1\cdot {\bf n}_j]^2=0
} \right.} \right)=1\,,\,\,\,\,\,\,
\forall\,{\bf n}_l \bot {\bf n}_k\,.
\label{tresprima}
\end{equation}
These two properties would play the same role as, respectively, 
properties (\ref{tresdos}) and (\ref{tres}). Thus
we can always find a set of inferences to build
a ladder proof. 
\end{sloppypar}

In fact, similar proofs exist for any maximally entangled state of two
spin-$s$ particles (with $s\ge 1$). Each of these states has infinite Schmidt 
decompositions and therefore there exists a prediction with certainty
between each local nondegenerate observable of one particle and
other local nondegenerate observable of the other particle. On the other hand,
there are local degenerate observables related with the previous 
nondegenerate observables, like $({\bf S}_1\cdot {\bf n}_j)^2$ is 
related with ${\bf S}_1\cdot {\bf n}_j$, which form an orthogonal
resolution of the identity of the Hilbert space of the corresponding
particle. This resolution of the identity would play the same role as relation
(\ref{resolution}) plays in the previous proof.

\section{On experiments}
Until now we have been reasoning with thought experiments. 
In this section, I will mention some of the advantages and disadvantages
of this ladder argument in a real experiment to test
local realism. 
Real experiments based on the ladder proof proposed in this paper will 
share some common features with experiments
based on Hardy's argument \cite{TBMM95,BDD97}
or on its ladder extension \cite{BBDH97}. For instance, 
since almost all the necessary experiments are measurements to confirm 
predictions with certainty, and since perfect certainties are hard to
find in a laboratory (see the results of \cite{BBDH97,TBMM95,BDD97}), 
some inequalities must be derived to deal with the data
\cite{BBDH97,TBMM95,BDD97,Mermin95}.
With this analysis, real experiments are not expected to elude 
the detection efficiency loophole \cite{GCS}, and therefore 
they will not provide more conclusive experimental tests against local realism
than previous tests of Bell's theorem \cite{GCS}.

On the other hand, the ladder proof proposed in this paper 
presents some advantages and disadvantages in
relation to the ladder proof by Hardy et al.
Pros: Maximally entangled states are easier to produce in a laboratory since
some of them are associated to a conserved quantity of a physical 
system after its decay into two parts. 
Since my proof works for all the pairs,
in principle, no postselection is needed. Only a finite number of observables
are needed.
Cons: I need at least a two-part three-level system. 
Each step of the ladder would require more experiments than in the 
case of Hardy et al. However, these experiments are always of
the same kind. They consist on
measuring the square of one spin component on one of the
particles and the square of the same spin component or one orthogonal to it 
of the other particle. On the contrary, in the experiment by
Hardy et al.\, the relative orientation of 
the polarizers changes in every step of the proof. More cons:
The geometrical argument requires measuring a triad of the square of
spin components in three mutually orthogonal directions
(or the equivalent observables if a different physical system is considered).
One way to do it is proposed in \cite{KS67}, however, it could be 
difficult to do this in practice. 

\begin{sloppypar}
\section{Ladder proofs of Bell's theorem versus proofs of the Kochen-Specker 
with locality theorem}
In this section I clarify the differences between ladder proofs of
Bell's theorem and proofs of the so-called
Kochen-Specker with locality (KSL) theorem. The KSL theorem shows 
that, for two spin-$1$ particles in the singlet state,
there is no hidden variables theory 
that satisfies separability, locality and some 
additional assumptions.
This result was first proved by Heywood and Redhead in 1983 \cite{HR83} 
and then reelaborated many times since \cite{varios}. 
Its proof is based on two points: 
First, on Kochen-Specker (KS) geometric proofs \cite{KS67}, which show 
that noncontextual values explaining
all quantum predictions are impossible for certain sets of observables of 
a single spin-$1$ particle
(usually these observables are squares of components of spin or other 
observables related to them, as in our ladder proof). Second, 
on EPR condition for elements of reality \cite{EPR35}.
But this condition is used here in a different way than in the ladder proofs: 
it is used to justify why in the singlet state of two spin-$1$ particles 
the previously mentioned observables must have a predefined value.
In contrast, ladder proofs use EPR condition to 
make predictions with certainty. 
 
Indeed, no published proof of the KSL theorem can be used to 
construct a ladder proof of Bell's theorem.  
To illustrate this point consider the following example.
Consider a singlet state of two spin-$1$ particles and on each of
them, consider
the simplest known KS geometric proof in a 
three dimensional Hilbert
space, due to Conway and Kochen \cite{Peres93}. 
All the directions used in the following explanation 
belong to this geometric proof.
Suppose that we make a measurement on the first particle and found that
$\left[ {{\bf S}_1\cdot (1,0,0)} \right]^2=0$. This implies that, for instance,
$\left[ {{\bf S}_2\cdot (0,1,-1)} \right]^2=\hbar ^2$. 
Then, this would imply that one of
$\left[ {{\bf S}_1\cdot (1,1,1)} \right]^2$ or
$\left[ {{\bf S}_1\cdot (-2,1,1)} \right]^2$ must be $0$ 
(and the other $\hbar ^2$). But which one? To decide it, one would
need to know the value in a direction orthogonal both to
$(0,1,-1)$ and $(1,1,1)$, or in a direction orthogonal to $(0,1,-1)$ and
$(-2,1,1)$, but such directions are not contained in the geometric proof 
by Conway and Kochen. Therefore, using this geometric proof
one cannot decide which one must be $0$. 
The same problem occurs sooner or 
later using every published geometric proof of
the KS theorem and therefore occurs in every proof of the KSL theorem.
In contrast, ladder proofs are based only on EPR inferences. 
Therefore, in a ladder
proof one must be able to predict with certainty all the 
outcomes involved in the proof, except the two at the beginning
($A_K$ and $B_K$ in the ladder proof by Hardy et al.,
or $A_{4K}$ and $B_{4K}$ in the proof proposed in this paper).
\end{sloppypar}

\section{Conclusions}
Hardy's argument \cite{Hardy93} is ``the best
version of Bell's theorem'' \cite{Mermin95} and possesses
``the highest attainable degree of simplicity 
and physical insight'' \cite{BDD97}. 
The recent ladder extension \cite{Hardy97,BBDH97} is
an improvement in the sense that a greater proportion of
the pairs is subject to a contradiction with local realism.
However, it does not work for maximally entangled states. 
In this paper, I have presented 
a proof which fills the most important holes left by the 
ladder extension by Hardy et al.: the new proof 
works for maximally entangled states 
of two spin-$s$ particles (with $s\ge 1$), 
and the proportion of the pairs subject
to a contradiction with local realism becomes $100\%$.
The experimental implementation of this proof could be 
achieved with present day technology, although in practice it would not provide
more conclusive results than previous tests of Bell's theorem.

\section{Acknowledgments}
The author thanks Guillermo Garc\'{\i}a Alcaine, 
Gonzalo Garc\'{\i}a de Polavieja, Lucien Hardy 
and Asher Peres for useful discussions and comments,
Jos\'{e} Luis Cereceda for drawing my attention to Ref.\ \cite{GK97},
and Carlos Serra for proofreading.

\pagebreak

\pagebreak

%Figura 1

\begin{figure}
\begin{picture}(120,160)(20,2)

%Figura (a)

\put(15,-20){\begin{picture}(120,120)

%Principio de las partes comunes a (a), (b) y (c)

\put(30,60){\circle*{5}}
\put(90,60){\circle*{5}}
\put(30,90){\circle*{5}}
\put(90,90){\circle*{5}}

\put(30,90){\line(2,-1){62}}
\put(30,90){\vector(2,-1){48}}
\put(90,90){\line(-2,-1){62}}
\put(90,90){\vector(-2,-1){48}}

\put(30,60){\line(1,0){60}}
\put(59,64){\line(0,-1){8}}
\put(61,64){\line(0,-1){8}}

\put(8,60){$A_0$}
\put(99,60){$B_0$}
\put(8,90){$A_1$}
\put(99,90){$B_1$}

%Final de las partes comunes a (a), (b) y (c)

\put(50,26){\Large (a)}
\put(55,98){$P_1$}
\put(30,90){\line(1,0){60}}

\end{picture}}

%Figura (b)

\put(145,-20){\begin{picture}(120,150)

%Principio de las partes comunes a (a), (b) y (c)

\put(30,60){\circle*{5}}
\put(90,60){\circle*{5}}
\put(30,90){\circle*{5}}
\put(90,90){\circle*{5}}

\put(30,90){\line(2,-1){62}}
\put(30,90){\vector(2,-1){48}}
\put(90,90){\line(-2,-1){62}}
\put(90,90){\vector(-2,-1){48}}

\put(30,60){\line(1,0){60}}
\put(59,64){\line(0,-1){8}}
\put(61,64){\line(0,-1){8}}

\put(8,60){$A_0$}
\put(99,60){$B_0$}
\put(8,90){$A_1$}
\put(99,90){$B_1$}

%Final de las partes comunes a (a), (b) y (c)

\put(50,26){\Large (b)}
\put(55,128){$P_2$}
\put(30,120){\line(1,0){60}}

\put(30,120){\circle*{5}}
\put(90,120){\circle*{5}}

\put(30,120){\line(2,-1){62}}
\put(90,120){\line(-2,-1){62}}
\put(30,120){\vector(2,-1){48}}
\put(90,120){\vector(-2,-1){48}}

\put(8,120){$A_2$}
\put(99,120){$B_2$}

\end{picture}}

%Figura (c)

\put(285,-20){\begin{picture}(120,190)

%Principio de las partes comunes a (a), (b) y (c)

\put(30,60){\circle*{5}}
\put(90,60){\circle*{5}}
\put(30,90){\circle*{5}}
\put(90,90){\circle*{5}}

\put(30,90){\line(2,-1){62}}
\put(30,90){\vector(2,-1){48}}
\put(90,90){\line(-2,-1){62}}
\put(90,90){\vector(-2,-1){48}}

\put(30,60){\line(1,0){60}}
\put(59,64){\line(0,-1){8}}
\put(61,64){\line(0,-1){8}}

\put(8,60){$A_0$}
\put(99,60){$B_0$}
\put(8,90){$A_1$}
\put(99,90){$B_1$}

%Final de las partes comunes a (a), (b) y (c)

\put(50,26){\Large (c)}
\put(55,188){$P_K$}
\put(30,180){\line(1,0){60}}

\multiput(30,120)(0,30){3}{\circle*{5}}
\multiput(90,120)(0,30){3}{\circle*{5}}

\multiput(30,180)(0,-30){2}{\line(2,-1){62}}
\multiput(30,180)(0,-30){2}{\vector(2,-1){48}}
\multiput(90,180)(0,-30){2}{\line(-2,-1){62}}
\multiput(90,180)(0,-30){2}{\vector(-2,-1){48}}

\put(6,180){$A_K$}
\put(-5,150){$A_{K-1}$}
\put(-5,120){$A_{K-2}$}

\put(99,180){$B_K$}
\put(99,150){$B_{K-1}$}
\put(99,120){$B_{K-2}$}

%Flechas cortadas

\put(30,120){\vector(2,-1){18}}
\put(90,120){\vector(-2,-1){18}}
\put(30,90){\line(2,1){20}}
\put(90,90){\line(-2,1){20}}

\put(50,100){\vector(-2,-1){8}}
\put(70,100){\vector(2,-1){8}}

\end{picture}}

\end{picture}

\caption{Diagrams for the ladder proofs for two spin-$\frac{1}{2}$ particles
by Hardy et al.
Original proof by Hardy with two observables on
each particle (a). 
Ladder proof with three observables (b) and 
ladder proof with $K+1$ observables (c).}
\vspace{3cm}
\end{figure}
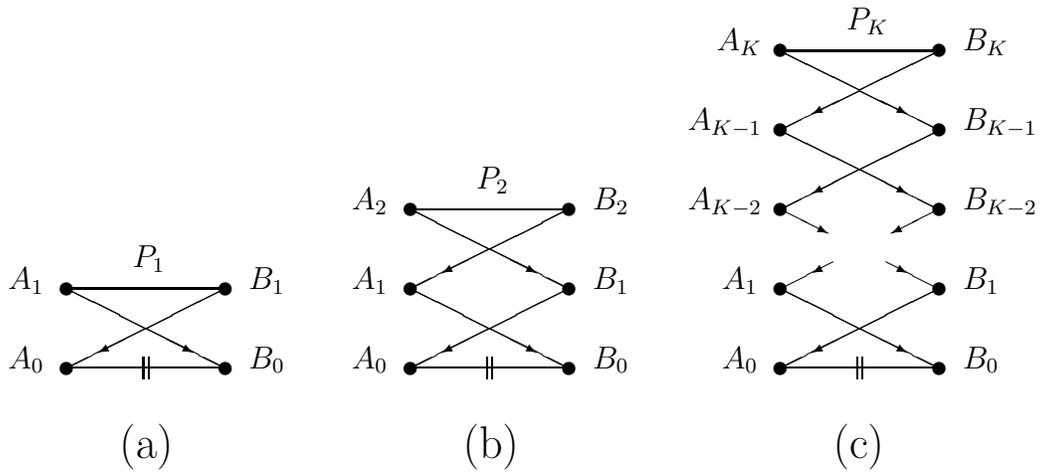

\pagebreak

%Figura 2

\begin{figure}
\begin{picture}(360,160)

%Diagrama

\put(134,-13){\begin{picture}(120,160)

\multiput(30,30)(0,30){5}{\circle*{5}}
\multiput(90,30)(0,30){5}{\circle*{5}}

\put(30,120){\line(1,-1){60}}
\put(30,120){\vector(1,-1){40}}
\put(90,120){\line(-1,-1){60}}
\put(90,120){\vector(-1,-1){40}}

\put(30,150){\line(1,0){60}}

\put(30,30){\line(1,0){60}}
\put(59,34){\line(0,-1){8}}
\put(61,34){\line(0,-1){8}}

\put(8,30){$A_0$}
\put(99,30){$B_0$}
\put(8,60){$A_1$}
\put(99,60){$B_1$}
\put(8,90){$A_2$}
\put(99,90){$B_2$}
\put(8,120){$A_3$}
\put(99,120){$B_3$}
\put(8,150){$A_4$}
\put(99,150){$B_4$}

\put(55,158){$P_4$}

\put(30,150){\line(1,-1){60}}
\put(30,150){\vector(1,-1){40}}
\put(90,150){\line(-1,-1){60}}
\put(90,150){\vector(-1,-1){40}}
\put(30,150){\line(2,-1){60}}
\put(30,150){\vector(2,-1){52}}
\put(90,150){\line(-2,-1){60}}
\put(90,150){\vector(-2,-1){52}}

\thicklines
\put(30,90){\line(1,-1){60}}
\put(30,90){\vector(1,-1){50}}
\put(90,90){\line(-1,-1){60}}
\put(90,90){\vector(-1,-1){50}}
\put(30,60){\line(2,-1){60}}
\put(30,60){\vector(2,-1){48}}
\put(90,60){\line(-2,-1){60}}
\put(90,60){\vector(-2,-1){48}}

\end{picture}}

%Observables de la particula 1

\put(-38,-13){\begin{picture}(0,160)
\put(0,150){$[{\bf S}_1\cdot (1 ,0 ,0)]^2=0$}
\put(0,120){$[{\bf S}_1\cdot (\tan \phi ,-1 , \cot \theta)]^2=\hbar ^2$}
\put(0,90){$[{\bf S}_1\cdot (\tan \phi ,-1 , -\cot \theta)]^2=\hbar ^2$}
\put(0,60){$[{\bf S}_1\cdot (0 ,\cos \theta ,-\sin \theta)]^2=\hbar ^2$}
\put(0,30){$[{\bf S}_1\cdot (\cot \phi \csc ^2\theta ,1,-\cot \theta )]^2=0$}

\end{picture}}

%Observables de la particula 2

\put(160,-13){\begin{picture}(120,160)
\put(99,150){$[{\bf S}_2\cdot (\cos \phi ,\sin \phi ,0)]^2=0$}
\put(99,120){$[{\bf S}_2\cdot (0 ,\cos \theta ,-\sin \theta)]^2=\hbar ^2$}
\put(99,90){$[{\bf S}_2\cdot (0 ,\cos \theta ,\sin \theta)]^2=\hbar ^2$}
\put(99,60){$[{\bf S}_2\cdot (\tan \phi ,-1 , \cot \theta)]^2=\hbar ^2$}
\put(99,30){$[{\bf S}_2\cdot (\cot \phi \csc ^2\theta ,1,\cot \theta )]^2=0$}

\end{picture}}

\put(-43,6){\framebox(178,148){}}
\put(254,6){\framebox(166,148){}}

\end{picture}
\caption{Diagram for the ladder proof for two spin-$1$ particles and 5 
observables on each particle (center). Corresponding inferences on 
the first particle (left) and on the second particle (right).}
\vspace{3cm}
\end{figure}
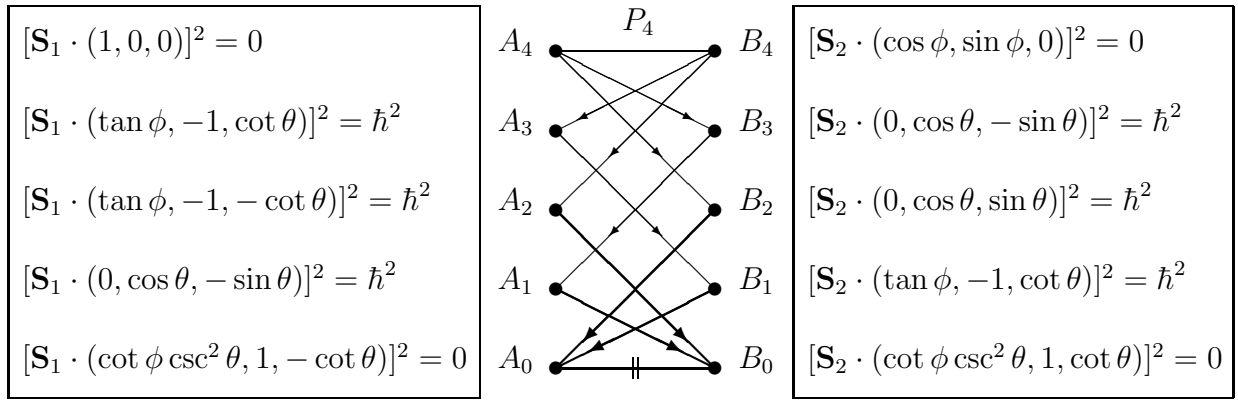

\pagebreak

%Figura 3

\begin{figure}
\begin{picture}(150,180)
\put(95,-200){\begin{picture}(150,180)
%\put(100,100){\vector(1,0){126}}
%\put(100,100){\vector(-1,-3){40}}
%\put(100,100){\vector(-1,3){40}}
%\thicklines
%\put(100,100){\vector(0,1){126}}
%\put(100,100){\vector(4,-3){107}}
%\put(100,100){\vector(-4,-3){107}}
%\put(-18,10){\bf i}
%\put(214,10){\bf j}
%\put(100,232){\bf k}
%\put(52,-32){\bf a}
%\put(233,100){\bf b}
%\put(54,226){\bf c}
\end{picture}}
\end{picture}
\vspace{8.75cm}
\caption{Relative orientations between the three orthogonal directions 
${\bf i}$, ${\bf j}$, ${\bf k}$ of the first particle and the 
three orthogonal directions ${\bf a}$, ${\bf b}$, ${\bf c}$ of the second
particle, corresponding to the geometrical argument of section~3.}
\end{figure}

\pagebreak

%Figura 4

\begin{figure}
\begin{picture}(360,340)

%Diagrama

\put(132,-133){\begin{picture}(120,520)

\multiput(30,30)(0,30){17}{\circle*{5}}
\multiput(90,30)(0,30){17}{\circle*{5}}

\put(30,120){\line(1,-1){60}}
\put(30,120){\vector(1,-1){40}}
\multiput(90,120)(0,30){5}{\line(-1,-1){60}}
\multiput(90,120)(0,30){5}{\vector(-1,-1){40}}
\multiput(90,330)(0,30){6}{\line(-1,-1){60}}
\multiput(90,330)(0,30){6}{\vector(-1,-1){40}}

\put(30,510){\line(1,0){60}}
\put(30,510){\line(2,-1){60}}
\put(30,510){\vector(2,-1){52}}
\put(30,510){\line(1,-1){60}}
\put(30,510){\vector(1,-1){40}}

\put(90,90){\circle{9}}
\put(90,120){\circle{9}}
\put(90,210){\circle{9}}
\put(90,240){\circle{9}}
\put(90,330){\circle{9}}
\put(90,360){\circle{9}}

%Flechas cortadas

\put(90,300){\vector(-1,-1){40}}
\put(90,270){\vector(-1,-1){16}}

\put(40,250){\vector(-1,-1){6}}
\put(40,250){\line(-1,-1){12}}
\put(62,242){\vector(-1,-1){11}}
\put(62,242){\line(-1,-1){32}}

%Fin de flechas cortadas

\put(90,510){\line(-2,-1){60}}
\put(90,510){\vector(-2,-1){52}}
\put(90,510){\line(-1,-1){60}}
\put(90,510){\vector(-1,-1){40}}

\put(30,30){\line(1,0){60}}
\put(59,34){\line(0,-1){8}}
\put(61,34){\line(0,-1){8}}

\put(2,30){$A_0$}
\put(105,30){$B_0$}
\put(2,60){$A_1$}
\put(105,60){$B_1$}
\put(2,90){$A_2$}
\put(105,90){$B_2$}
\put(2,120){$A_3$}
\put(105,120){$B_3$}
\put(2,150){$A_4$}
\put(105,150){$B_4$}
\put(2,180){$A_5$}
\put(105,180){$B_5$}
\put(2,210){$A_6$}
\put(105,210){$B_6$}
\put(2,240){$A_7$}
\put(105,240){$B_7$}
\put(-14,270){$A_{4(K-2)}$}
\put(96,270){$B_{4(K-2)}$}
\put(-9,300){$A_{4K-7}$}
\put(98,300){$B_{4K-7}$}
\put(-9,330){$A_{4K-6}$}
\put(98,330){$B_{4K-6}$}
\put(-9,360){$A_{4K-5}$}
\put(98,360){$B_{4K-5}$}
\put(-14,390){$A_{4(K-1)}$}
\put(96,390){$B_{4(K-1)}$}
\put(-9,420){$A_{4K-3}$}
\put(98,420){$B_{4K-3}$}
\put(-9,450){$A_{4K-2}$}
\put(98,450){$B_{4K-2}$}
\put(-9,480){$A_{4K-1}$}
\put(98,480){$B_{4K-1}$}
\put(0,510){$A_{4K}$}
\put(100,510){$B_{4K}$}

\put(52,518){$P_{4K}$}

\thicklines
\put(30,90){\line(1,-1){60}}
\put(30,90){\vector(1,-1){50}}
\put(90,90){\line(-1,-1){60}}
\put(90,90){\vector(-1,-1){50}}
\put(30,60){\line(2,-1){60}}
\put(30,60){\vector(2,-1){48}}
\put(90,60){\line(-2,-1){60}}
\put(90,60){\vector(-2,-1){48}}
\put(30,210){\line(1,-1){60}}
\put(30,210){\vector(1,-1){40}}
\put(30,240){\line(1,-1){60}}
\put(30,240){\vector(1,-1){40}}
\put(30,330){\line(1,-1){60}}
\put(30,330){\vector(1,-1){40}}
\put(30,360){\line(1,-1){60}}
\put(30,360){\vector(1,-1){40}}
\put(30,450){\line(1,-1){60}}
\put(30,450){\vector(1,-1){40}}
\put(30,480){\line(1,-1){60}}
\put(30,480){\vector(1,-1){40}}
\put(30,150){\line(1,0){60}}
\put(30,150){\vector(1,0){50}}
\put(30,180){\line(1,0){60}}
\put(30,180){\vector(1,0){50}}
\put(30,270){\line(1,0){60}}
\put(30,270){\vector(1,0){50}}
\put(30,300){\line(1,0){60}}
\put(30,300){\vector(1,0){50}}
\put(30,390){\line(1,0){60}}
\put(30,390){\vector(1,0){50}}
\put(30,420){\line(1,0){60}}
\put(30,420){\vector(1,0){50}}

\end{picture}}

%Observables de la particula 1

\put(-48,-133){\begin{picture}(0,160)
\put(0,510){$[{\bf S}_1\cdot (1 ,0 , 0)]^2=0$}
\put(0,480){$[{\bf S}_1\cdot (\tan \phi ,-1 , \cot \theta_{1})]^2=\hbar ^2$}
\put(0,450){$[{\bf S}_1\cdot (\tan \phi ,-1 , -\cot \theta_{1})]^2=\hbar ^2$}
\put(0,420){$[{\bf S}_1\cdot (0 ,\cos \theta_{1} ,\sin \theta_{1})]^2=\hbar ^2$}
\put(0,390){$[{\bf S}_1\cdot (0 ,\cos \theta_{1} ,-\sin \theta_{1})]^2=\hbar ^2$}

\put(0,360){$[{\bf S}_1\cdot (\tan \phi ,-c_{2}^{-1}\sin \theta _{2} ,$}
\put(72,345){$c_{2}^{-1}\cos \theta _{2})]^2=\hbar ^2$}

\put(0,330){$[{\bf S}_1\cdot (\tan \phi ,-c_{2}^{-1}\sin \theta _{2} ,$}
\put(63,315){$-c_{2}^{-1}\cos \theta _{2})]^2=\hbar ^2$}

\put(0,300){$[{\bf S}_1\cdot (0 ,\cos \theta_{2} ,\sin \theta_{2})]^2=\hbar ^2$}
\put(0,270){$[{\bf S}_1\cdot (0 ,\cos \theta_{2} ,-\sin \theta_{2})]^2=\hbar ^2$}

\put(0,240){$[{\bf S}_1\cdot (\tan \phi ,-c_{K-1}^{-1}\sin \theta _{K-1} ,$}
\put(50,225){$c_{K-1}^{-1}\cos \theta _{K-1})]^2=\hbar ^2$}

\put(0,210){$[{\bf S}_1\cdot (\tan \phi ,-c_{K-1}^{-1}\sin \theta _{K-1} ,$}
\put(40,195){$-c_{K-1}^{-1}\cos \theta _{K-1})]^2=\hbar ^2$}

\put(0,180){$[{\bf S}_1\cdot (0 ,\cos \theta_{K-1} ,$}
\put(77,165){$\sin \theta_{K-1})]^2=\hbar ^2$}

\put(0,150){$[{\bf S}_1\cdot (0 ,\cos \theta_{K-1} ,$}
\put(66,135){$-\sin \theta_{K-1})]^2=\hbar ^2$}

\put(0,120){$[{\bf S}_1\cdot (\tan \phi ,-c_{K}^{-1}\sin \theta _{K} ,$}
\put(68,105){$c_{K}^{-1}\cos \theta _{K})]^2=\hbar ^2$}

\put(0,90){$[{\bf S}_1\cdot (\tan \phi ,-c_{K}^{-1}\sin \theta _{K} ,$}
\put(59,75){$-c_{K}^{-1}\cos \theta _{K})]^2=\hbar ^2$}

\put(0,60){$[{\bf S}_1\cdot (0 ,\cos \theta_{K} ,-\sin \theta_{K})]^2=\hbar ^2$}

\put(0,30){$[{\bf S}_1\cdot (\cot \phi ,c_{K}\sin \theta _{K} ,$}
\put(63,15){$-c_{K}\cos \theta _{K})]^2=0$}

\end{picture}}

%Observables de la particula 2

\put(176,-133){\begin{picture}(120,160)
\put(99,510){$[{\bf S}_2\cdot (\cos \phi ,\sin \phi ,0)]^2=0$}
\put(99,480){$[{\bf S}_2\cdot (0 ,\cos \theta_{1} ,\sin \theta_{1})]^2=\hbar ^2$}
\put(99,450){$[{\bf S}_2\cdot (0 ,\cos \theta_{1} ,-\sin \theta_{1})]^2=\hbar ^2$}

\put(99,420){$[{\bf S}_2\cdot (\cot \phi ,c_{1}\sin \theta _{1} ,$}
\put(167,405){$-c_{1}\cos \theta _{1})]^2=0$}

\put(99,390){$[{\bf S}_2\cdot (\cot \phi ,c_{1}\sin \theta _{1} ,$}
\put(177,375){$c_{1}\cos \theta _{1})]^2=0$}

\put(99,360){$[{\bf S}_2\cdot (0 ,\cos \theta_{2} ,\sin \theta_{2})]^2=\hbar ^2$}
\put(99,330){$[{\bf S}_2\cdot (0 ,\cos \theta_{2} ,-\sin \theta_{2})]^2=\hbar ^2$}

\put(99,300){$[{\bf S}_2\cdot (\cot \phi ,c_{2}\sin \theta _{2} ,$}
\put(168,285){$-c_{2}\cos \theta _{2})]^2=0$}

\put(99,270){$[{\bf S}_2\cdot (\cot \phi ,c_{2}\sin \theta _{2} ,$}
\put(177,255){$c_{2}\cos \theta _{2})]^2=0$}

\put(99,240){$[{\bf S}_2\cdot (0 ,\cos \theta_{K-1} ,$}
\put(175,225){$\sin \theta_{K-1})]^2=\hbar ^2$}

\put(99,210){$[{\bf S}_2\cdot (0 ,\cos \theta_{K-1} ,$}
\put(165,195){$-\sin \theta_{K-1})]^2=\hbar ^2$}

\put(99,180){$[{\bf S}_2\cdot (\cot \phi ,c_{K-1}\sin \theta _{K-1} ,$}
\put(139,165){$-c_{K-1}\cos \theta _{K-1})]^2=0$}

\put(99,150){$[{\bf S}_2\cdot (\cot \phi ,c_{K-1}\sin \theta _{K-1} ,$}
\put(149,135){$c_{K-1}\cos \theta _{K-1})]^2=0$}

\put(99,120){$[{\bf S}_2\cdot (0 ,\cos \theta_{K} ,-\sin \theta_{K})]^2=\hbar ^2$}
\put(99,90){$[{\bf S}_2\cdot (0 ,\cos \theta_{K} ,\sin \theta_{K})]^2=\hbar ^2$}

\put(99,60){$[{\bf S}_2\cdot (\tan \phi ,-c_{K}^{-1}\sin \theta _{K} ,$}
\put(167,45){$c_{K}^{-1}\cos \theta _{K})]^2=\hbar ^2$}

\put(99,30){$[{\bf S}_2\cdot (\cot \phi ,c_{K}\sin \theta _{K} ,$}
\put(171,15){$c_{K}\cos \theta _{K})]^2=0$}

\end{picture}}

\put(-52,-125){\framebox(166,519){}}
\put(271,-125){\framebox(166,519){}}

\end{picture}
\vspace{4.4cm}
\caption{Diagram for the chain of predictions between $4K+1$
observables on each particle (center) used in section~4. 
Explicit predictions on 
the first particle (left) and on the second particle (right).}
\end{figure}
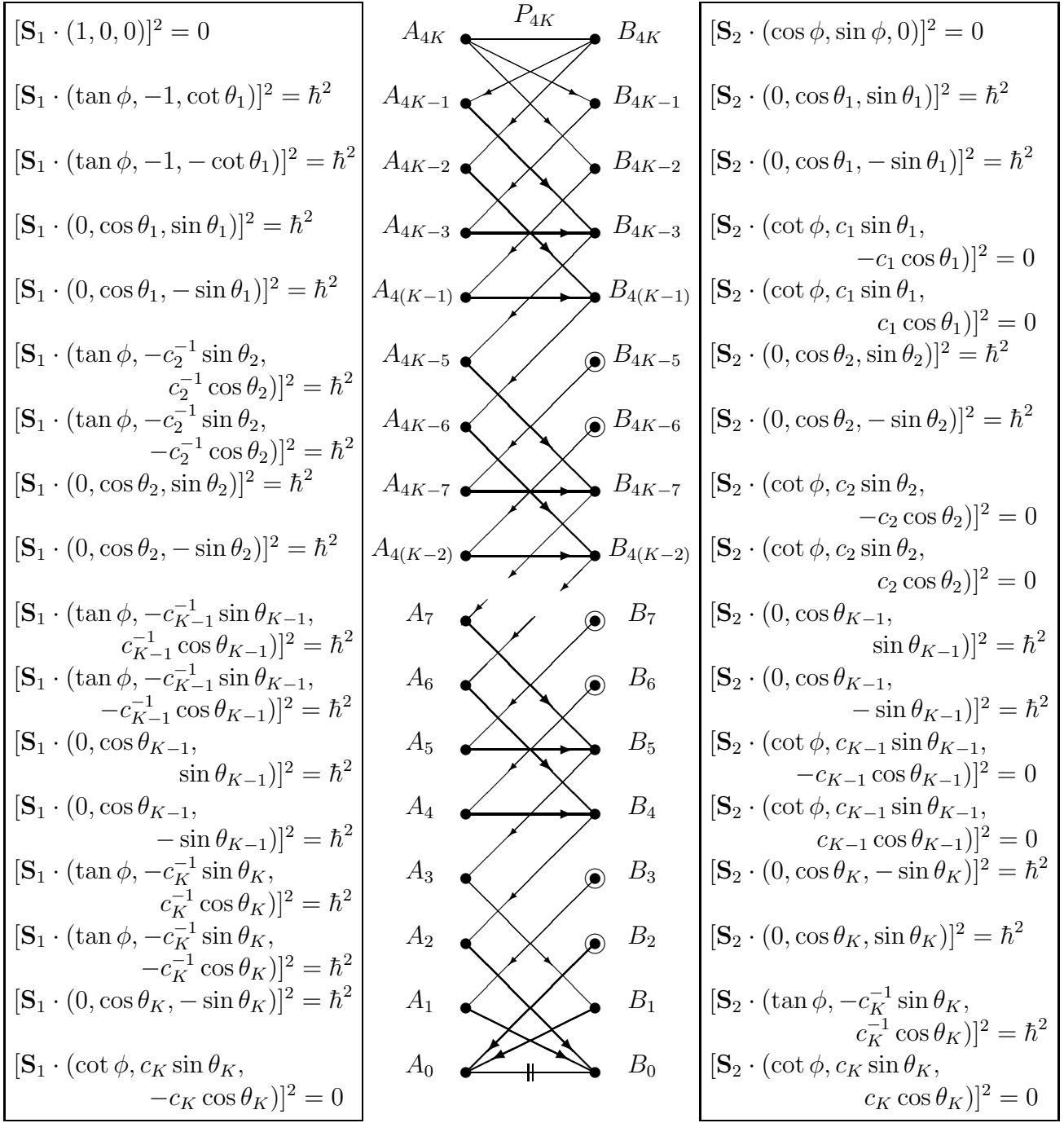

\pagebreak

%Figura 5

\begin{figure}
\begin{picture}(150,180)
\put(95,-200){\begin{picture}(150,180)
%\put(100,100){\vector(1,-3){40}}
%\put(100,100){\vector(-3,2){102}}
%\put(100,100){\vector(4,3){100}}
%\thicklines
%\put(100,100){\vector(0,1){126}}
%\put(100,100){\vector(4,-1){120}}
%\put(100,100){\vector(-4,-3){102}}
%\put(-12,14){\bf i}
%\put(226,62){\bf j}
%\put(100,232){\bf k}
%\put(138,-32){\bf a}
%\put(206,175){\bf b}
%\put(-13,171){\bf c}
\end{picture}}
\end{picture}
\vspace{8.75cm}
\caption{Relative orientations between the three orthogonal directions 
${\bf i}$, ${\bf j}$, ${\bf k}$ of the first particle and the 
three orthogonal directions ${\bf a}$, ${\bf b}$, ${\bf c}$ of the second
particle, corresponding to the geometrical argument of section~4.}
\end{figure}

\end{document}